\documentclass{article}

\usepackage{arxiv}

\usepackage[utf8]{inputenc} 
\usepackage[T1]{fontenc}    
\usepackage{hyperref}       
\usepackage{url}            
\usepackage{booktabs}       
\usepackage{amsfonts}       
\usepackage{nicefrac}       
\usepackage{microtype}      
\usepackage{lipsum}		
\usepackage{authblk}    
\usepackage{graphicx}
\usepackage{doi}

\title{\emph{Nondestructive KPFM-assisted Quality Control in Fabrication of GaAs High-Speed Electronics}}

\author[1,2,$\star$]{Alexander Shurakov}
\author[1]{Natalia Kaurova}
\author[1,2]{Ivan Belikov}
\author[3]{Tatyana Zilberley}
\author[1,2]{Anatoliy Prikhodko}
\author[1]{Boris Voronov}
\author[1,2]{Gregory Gol{'}tsman}

\affil[1]{Moscow Pedagogical State University, Moscow, 119991, Russia}
\affil[2]{National Research University Higher School of Economics, Moscow, 101000, Russia}
\affil[3]{Space Research Institute of the Russian Academy of Sciences, Moscow, 117997, Russia}
\affil[$\star$]{email: alexander@rplab.ru}

\renewcommand{\shorttitle}{Shurakov et al.}

\begin{document}
\maketitle

\begin{abstract}
In this paper, we report on the method of nondestructive quality control that can be used in fabrication of GaAs high-speed electronics. The method relies on the surface potential mapping and enables rigid \textit{in vivo} analysis of transport properties of an active electronic device incorporated into a complex integrated circuit. The study is inspired by our ongoing development of a millimeter wave intelligent reflective surface for 6G communications. To provide desired beamforming capabilities, such a surface should utilize hundreds of identical microscale GaAs diode switches with series resistance of a few ohms. Thus, we develop a ladder-like layered ohmic contact to heavily Si-doped GaAs and cross-study it via transmission line method and Kelvin probe force microscopy. The contact resistivity as low as 0.15~$\mu \Omega \,$cm$^2$ is measured resulting in only a 0.6~$\Omega$ of resistance for the contact area of 3$\times$3~$\mu$m$^2$. Moreover, the tendencies observed suggest that one can rigidly analyze the evolution of contact resistance and the profile of resistivity under contact in response to rapid thermal annealing, once the surface potential map across the ``ladder'' is known.
\end{abstract}


\section{Introduction}
Nowadays, GaAs is in vast demand in the monolithic microwave integrated circuit (MMIC) receiver technology including development of power amplifiers, frequency multipliers, coherent and direct detectors for both scientific and civilian purposes \cite{mehdi2017thz,tokumitsu2014application}.

In this paper, we report on the method of nondestructive quality control that can be used in fabrication of GaAs high-speed electronics. The method relies on the surface potential mapping and enables rigid \textit{in vivo} analysis of transport properties of an active electronic device incorporated into a complex integrated circuit (IC). The study is inspired by our ongoing development of a millimeter wave (mmWave) intelligent reflective surface (IRS), which is currently among actively developing technologies for 6G communication networks \cite{A3,A4,A5,A6}. IRS is meant to significantly improve the quality of mmWave communication in the case of blocking the line of sight between the transmitter and receiver \cite{A7}. The use of IRS can significantly reduce the power consumption during data transmission compared to using the technology of communication with multiple-input and multiple-output (MIMO). In general, IRS is an array of electrically controlled elements used to locally adjust the phase of the incident wave. In the case of a 1-bit phase shift resolution for wave reflection, when each IRS element has only two states +0$^\circ$ and +180$^\circ$, a noticeable decrease in the reflected beam directivity can be observed due to the presence of significant phase errors \cite{A8,A9}. Increasing the phase resolution improves the performance of IRS, but increases its complexity and the fabrication cost. The optimal combination of price and quality for IRS is achieved when it utilizes elements with 2-bit phase resolution \cite{A8,A10}. To date, there are only a few works on IRS with 2-bit elements \cite{A10,A11,A12,A13} in the literature, and there are no works on these devices with operating frequencies above 30 GHz. However, the existing information hints that, to provide desired beamforming capabilities at frequencies above 100 GHz, IRS should utilize hundreds of identical microscale GaAs diode switches with series resistance of a few ohms.

Implementation of an ohmic contact to semiconductor is a crucial part of fabrication technology that largely impacts on the overall performance of the device being fabricated. The transmission line method (TLM) is widely used to characterize properties of an ohmic contact to a planar semiconductor structure \cite{berger1972models}. Staying simple to use, TLM yet has certain drawbacks originating from the assumption of the uniform current distribution within a transmission line (TL) and provides result with certain error, which is geometry dependent \cite{grover2016effect}. In addition to solely measuring the contact resistance value, the grooved transmission line method (GTLM) \cite{heiblum1982characteristics} enables evaluation of the profile of resistivity under contact. This option is useful if the dissolution of semiconductor in melted metallization system with further recrystallization takes place due to rapid thermal annealing (RTA).

In this work, we develop a ladder-like layered ohmic contact to heavily Si-doped GaAs and cross-study it via GTLM and Kelvin probe force microscopy (KPFM). The contact resistivity as low as 0.15~$\mu \Omega \,$cm$^2$ is measured resulting in only a 0.6~$\Omega$ of resistance for the contact area of 3$\times$3~$\mu$m$^2$. Moreover, the tendencies observed suggest that one can rigidly analyze the evolution of contact resistance and the profile of resistivity under contact in response to RTA, once the surface potential map across the ``ladder'' is known. While fabricating IRS, surface potential mapping of the ``ladders'' in array can be employed to analyze transport properties of its elements locally. Thus, fabrication process can be tuned to achieve high homogeneity of the IRS elements along with their decent individual performance. More details on the outcome of our studies are presented further in the text. 

\section{Fabrication of experimental samples}
\begin{figure}[!b]
\centering
\includegraphics[width=0.5\textwidth]{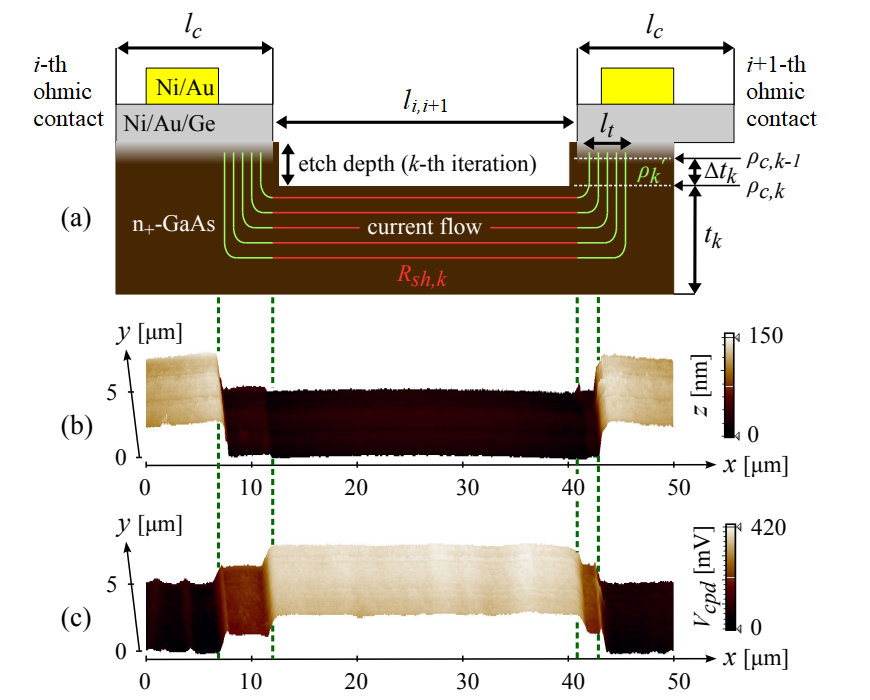}
\caption{(a) Schematic diagram (side view) of the TL section confined by the first pair of ohmic contacts. Figure also contains KPFM images of the relief (b) and the surface potential (c) of the TL section obtained within a dual-pass operating regime of the microscope.}
\label{fig1}
\end{figure}

Basic fabrication routine is similar to that previously employed by us in \cite{shurakov2018ti}. Yet, we choose the ohmic contact metallization system based on the Ni/Au/Ge/Ni/Au (5/45/20/5/100 nm, Au is the topmost layer) sandwich, which is deposited on a 2 $\mu$m thick Si-doped GaAs by the means of e-beam evaporation within two iterations. The dopant concentration is equal to $5\times10^{18}$ cm$^{-3}$, the handle wafer presented by SI-GaAs has a thickness of 350 $\mu$m. As shown in figure \ref{fig1}(a), this revised routine is employed to fabricate a series of TL samples utilizing ladder-like ohmic contact structures with an offset between Ni/Au/Ge and Ni/Au stacks of a few microns. Such a geometry of the contacts is developed to provide the possibility of measuring the Ge and n$^+$-GaAs surface potentials relative to the Au zero level. At the final stage of fabrication, the samples are exposed to a heat treatment via rapid thermal annealing. While sweeping the annealing temperature ($T_{rta}$) from 300 to 450$^{\circ}$C, we fix the nitrogen flow rate and duration of the annealing process to 1350 l/h and 60 s, respectively. Geometry of the fabricated TL samples is detailed in table \ref{tab:geom}.

\begin{table}[t!]
\centering
\caption{Geometry of TL samples. Here $w_{tl}$ and $w_c$ are the widths of TL and its ohmic contacts, respectively.}
\label{tab:geom}
\setlength{\tabcolsep}{18pt}
\begin{tabular}{ccccc}
\hline 
$w_{tl}$ [$\mu$m] & $w_c$ [$\mu$m] & $l_c$ [$\mu$m] & $l_{i,i+1}$ [$\mu$m] & $i$\\
\hline
310 & 310 & 100 & $30 i$ & 1..9\\
\hline
\end{tabular}
\\
\footnotesize
\vspace{6pt}
\end{table}
\normalsize

\section{Research methods and results}
Once fabricated, TL samples are further studied by the means of KPFM. Figure \ref{fig1}(a) contains schematic diagram of the TL section confined by a pair of neighboring ohmic contacts. Outcome of the KPFM scans performed on the section’s top surface is also provided in the figure. As one can clearly see, the Au–Ge–n$^+$-GaAs transitions corresponding to the intervals $x\in(8, 12)$ and $x\in(41, 43)$ depicted by the green dashed lines are more pronounced in case of the surface potential (figure \ref{fig1}(c)) compared to that of the relief map (figure \ref{fig1}(b)). This fact hints that the surface potential mapping is a promising tool for analysis of the RTA impact. Before independent use, however, it needs to be performed together with a well-established analysis method such as TLM or, more preferably, GTLM.

\subsection{KPFM measurements}
During the measurement, the microscope is operated in intermediate mode \cite{melitz2011kelvin}, i.e., the cantilever tip does not physically touch surface of the sample. The cantilever is externally excited to oscillate by AC signal of certain amplitude ($V_{ac} + V_{dc}$) and frequency ($\omega$). Amplitude of the resulted oscillation is a function of spacing between the tip and the sample. Therefore, it is used as a feedback parameter to implement imaging. Contact potential difference (CPD) between the cantilever tip and the sample surface is measured as a constant voltage on the tip (V$_{cpd}$) needed to nullify the oscillation at fundamental frequency of the AC signal. Indeed, the component of the tip-sample electrostatic force causing the oscillation is defined as

\begin{equation}
\label{eq:F}
F_\omega = -{\partial C(z)}/{\partial z} \, (V_{dc} \pm V_{cpd}) \, V_{ac} \, sin(\omega t),
\end{equation}
where $\partial C(z)/ \partial z$ is the capacitance gradient between the tip and the sample along the direction perpendicular to the sample surface.

In our studies, we use commercially available modular scanning probe microscopy system NTEGRA \cite{ntegra} equipped with a high resolution high accuracy cantilevers HA NC/Pt \cite{hactlv} well-suited for KPFM measurements. The microscope is operating in a dual-pass regime, when mapping of the relief and surface potential is carried out within two back-to-back passes of a single scan. KPFM images are obtained with a spatial resolution of $\sim$100 nm in the $xy$-plane. The third coordinate resolution is optimized for each particular sample through manual tuning of the feed back loop gain in the range of 0.4-0.7 by Ziegler-Nichols method. The peak-to-peak fluctuations of the altitude and surface potential values within a scan due to the system noise are equal to 5 nm and 15 meV, respectively. The profiles are obtained as an average of 50 scans reducing the fluctuations to 0.7 nm (altitude) and 2.5 meV (surface potential).

\begin{figure}[!t]
\centering
\includegraphics[width=0.5\textwidth]{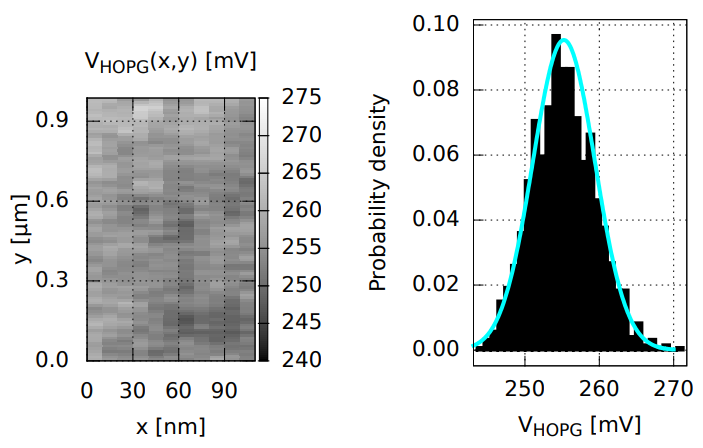}
\caption{The CPD voltage map (left panel) and probability density (right panel, bars) for the HOPG sample surface fitted with the normal distribution equation (right panel, solid line).}
\label{fig5}
\end{figure}

\begin{figure}[!b]
\centering
\includegraphics[width=1\textwidth]{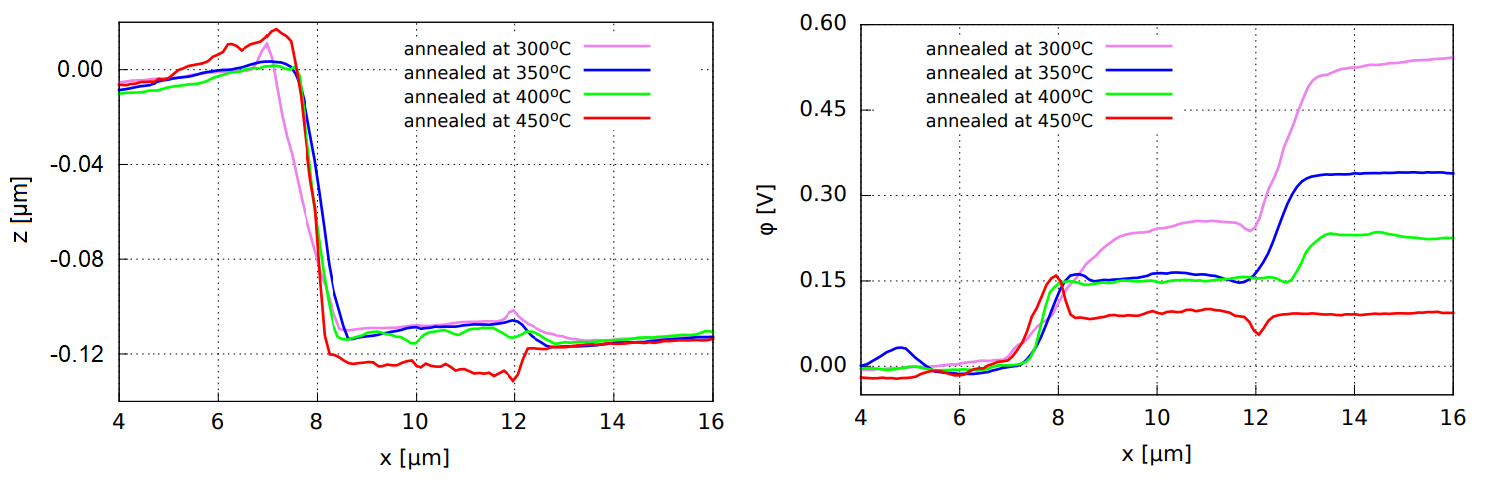}
\caption{Impact of RTA on the relief (left panel) and the relative surface potential (right panel) of the TL section in vicinity of the ohmic contact containing a ladder-like Au–Ge–n$^+$-GaAs transition on the top surface.}
\label{fig3and4}
\end{figure}

To evaluate absolute values of the surface potential, the cantilever tip work function ($\phi_{ct}$) is calibrated with the aid of Highly Oriented Pyrolytic Graphite (HOPG) possessing mosaic spread of 0.8--1.2$^{\circ}$ and work function of $4.475 \pm 0.005$~eV \cite{hansen2001standard}. The calibration is conducted in air at relative humidity of 52--55\%, a set lift-height from the sample surface of 10 nm and $V_{ac} = 10$~mV. Referring to figure \ref{fig5}, we measure the CPD voltage map for a $0.1 \times 1$~$\mu$m$^2$ surface area of the HOPG sample ($V_{HOPG}(x,y)$) and observed normal distribution with mean value of 255 mV and standard deviation of 4 mV for the voltages measured. Applying the tree-sigma rule to the $V_{HOPG}$ data, one can obtain $\phi_{ct} = 4.730 \pm 0.017$ eV calculated within a nonoscillating cantilever approximation described by equation \ref{eq:F}. However, it has to be borne in mind that the experiment conditions should result in a $-109 \pm 7$~meV offset of the CPD voltage value compared to that measured in vacuum \cite{liscio2008tip}. We also observe minor gradual decrease of the tip work function during numerous iterations of the calibration procedure. Therefore, all the curves further appeared in figure \ref{fig3and4} are provided with respect to the Au zero level.

Left panel of figure \ref{fig3and4} provides the relief profiles in vicinity of the Au–Ge–n$^+$-GaAs transition for the TL samples annealed at different temperatures. It is clearly seen from the figure that drastic difference appears between the curves corresponding to the annealing temperatures of 300–-400$^{\circ}$C and that of 450$^{\circ}$C, when the Ni/Au/Ge stack is no longer above the n$^+$-GaAs surface. Moreover, even without a rigid grain analysis, one can notice waviness of the relief profile within the Ge section for the TL sample annealed at 400$^{\circ}$C. We consider that as a suggestion of initiation of the semiconductor dissolution in metal followed by recrystallization due to the heat treatment \cite{blank2007mechanisms}. Indeed, the latter temperature is close to that of the AuGe alloy eutectic point which equals 356$^{\circ}$C \cite{kisiel2009attachment}.

Potential penetration of the Ni/Au/Ge stack into the  GaAs volume laying underneath and further redistribution of the melt is also suggested by the KPFM images presented in right panel of figure \ref{fig3and4}. The family of the surface potential profiles provided in the figure yields gradual decrease of the difference of potentials of the Ge and n$^+$-GaAs top surfaces in response to increasing $T_{rta}$. And the surface potential difference vanishes, when $T_{rta} = 450^{\circ}$C. It is necessary to note that the spike observed in figure \ref{fig3and4} in vicinity of $x = 8$~$\mu$m is due to the presence of edging appeared because of the photolithography process peculiarity. The results obtained partly agree with those presented in \cite{boumenou2018effect}: the n$^+$-GaAs surface potential of the sample annealed at 350$^{\circ}$C is also equal to $\sim$150 meV measured relative to the SI-GaAs level. We, however, observe noticeable dependence of the surface properties on heat treatment \cite{moison1987surface}.

Given all the facts mentioned above, we consider the TL samples annealed at 350$^{\circ}$C (cold sample) and 450$^{\circ}$C (hot sample) to be of interest for GTLM studies.

\subsection{GTLM studies}
As GTLM suggests, we conduct a series of successive measurements of the resistance values ($R_{i,i+1}$) of the TL sections confined by pairs of neighboring ohmic contacts. The measurements are carried out with the aid of a source-meter operating in a four-wire regime to eliminate parasitic impact of the probe-to-sample resistance. The resistance values obtained within the experiment obey a linear dependence on the distance separating the ohmic contacts between which they are measured. Assuming that sheet resistance of n$^+$-GaAs ($R_{sh}$) is identical in all the regions of TL, we express this dependence as

\begin{equation}
\label{eq:Ri,i+1}
R_{i,i+1} = R_{sh} \, w_c^{-1} \, (l_{i,i+1} + 2l_t). 
\end{equation}
Thus, the slope of the function defined by equation \ref{eq:Ri,i+1} provides $R_{sh}$ if $w_c$ is known. The corresponding transfer length ($l_t$) is extracted as the intersection point of the extrapolated $R_{i,i+1}(l_{i,i+1}$) curve with the abscissa axis multiplied by a factor of $-0.5$. Our measurements yield decrease of $l_t$ from 52.8 to 1.5 $\mu$m in response to sweeping $T_{rta}$ from 300 to 400$^{\circ}$C. And no noticeable difference is observed for the samples annealed at 400 and 450$^{\circ}$C.

\begin{figure}[!b]
\centering
\includegraphics[width=0.5\textwidth]{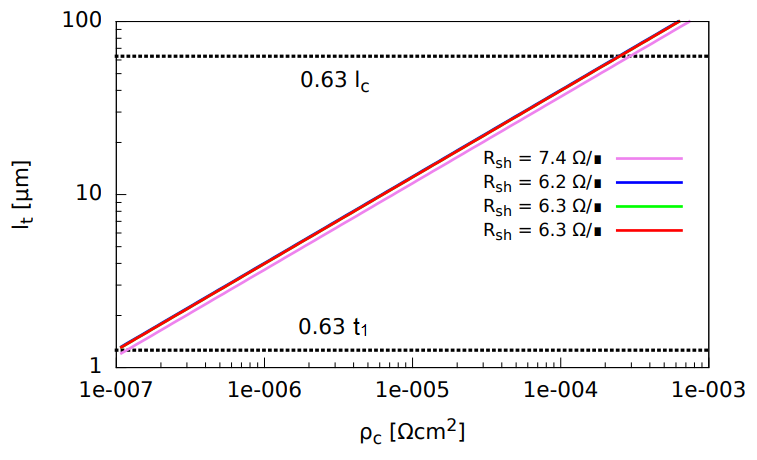}
\caption{Transfer length as a function of contact resistivity. Both axis are given in a logarithmic scale. The coloring is identical to that provided in figure \ref{fig3and4}.}
\label{fig2}
\end{figure}

Contact resistivities ($\rho_c$) of 2.7 and 0.15 $\mu \Omega$cm$^2$ are further calculated with the aid of equation \ref{eq:rhoc} for the cold and the hot TL samples, respectively. Thus, significantly lower contact resistivity is observed compared to that reported in the literature for the optimized AuGe/Ni/Au ohmic contacts to n$^+$-GaAs \cite{baranska2013ohmic}. We also would like to stress that all the measured values does not leave the trusted range defined by the TL geometry. Indeed, based on the comparison of the functional dependencies $l_t(\rho_c)$ obtained within the finite element and the transmission line methods, it is demonstrated that the latter provides no deviation of the $l_t$ value from that intrinsic to the TL structure if the value stays within the range from $0.63 t_1$ to $0.63 l_c$ \cite{grover2016effect}. Figure \ref{fig2} contains the $l_t = \sqrt{\rho_c \, R_{sh}^{-1}}$ plots for all the studied TL samples. As one can see from the figure, the sample geometry enables to accurately measure contact resistivities from $1.2\times10^{-7}$ to $2.5\times10^{-4}$ $\Omega$cm$^2$. 

\begin{figure}[!t]
\centering
\includegraphics[width=1\textwidth]{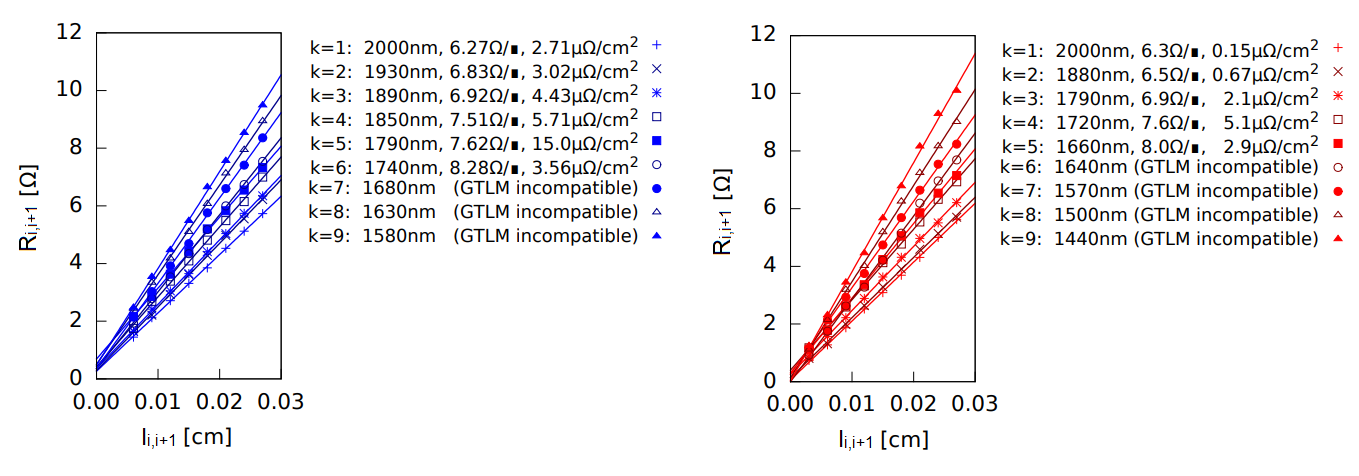}
\caption{Resistance of the TL section as a function of spacing between its ohmic contacts. The $R_{i,i+1} (l_{i,i+1})$ family is obtained for the samples annealed at 350$^{\circ}$C (left panel) and 450$^{\circ}$C (right panel). Each iteration of etching ($k = 1..6$) is denoted with a triple $t_k,\,R_{sh,k},\,\rho_{c,k}$.}
\label{fig6and7}
\end{figure}

\begin{figure}[!t]
\centering
\includegraphics[width=0.5\textwidth]{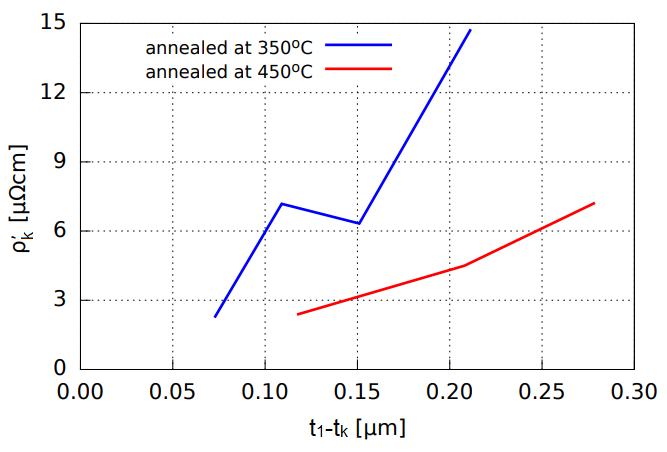}
\caption{Resistivity under ohmic contact as a function of immersion depth.}
\label{fig8}
\end{figure}

Upon successive etching of n$^+$-GaAs between all the ohmic contacts of TL down to thickness ($t_k$) comparable to that of the unetched TL ($t_1$) in terms of not disturbing the current distribution, we evaluate the resistivity of $k$-th layer under the ohmic contact surface ($\rho'_k$), i.e., at depth $t_1 - t_k$ and with thickness $\Delta t_k = |t_k - t_{k-1}|$, in accordance with equation \ref{eq:rhok}.

\begin{equation}
\label{eq:rhoc}
\rho_c = R_{sh} \, l^2_t.
\end{equation}

\begin{equation}
\label{eq:rhok}
\rho'_k = \sqrt{R_{sh,k} \, t_k \, \Delta t_k^{-1} \, (\rho_{c,k} - \rho_{c,k-1})}.
\end{equation}
During each iteration of the measurement, we etch the TL samples in H$_2$O:H$_2$O$_2$:Ammonia solution 25\% (80:8:2) for $\sim$1.5 s. Referring to \ref{fig6and7}, gradual increase of the cold sample contact resistivity is observed until the 6-th iteration of etching, when GTLM becomes inapplicable due to significant disturbance of the current distribution in TL. Such a situation is characterized by the appearance of negative difference of the $\rho_c$ values obtained within two successive iterations. The figure also shows that the same situation takes place during the 5-th iteration of etching for the hot sample. The profiles of resistivity under ohmic contact provided in \ref{fig8} suggest that annealing at 450$^{\circ}$C is significantly more efficient in case of thick semiconductor. Indeed, the ratio of the cold and hot sample resistivities is equal to 2:1 and 3.3:1 at distances of 150 and 210 nm below the ohmic contact surface, respectively; and the ratio keeps rising with further deepening.

\subsection{GaAs diode switches}
The results of our studies of the ohmic contact metillization system are further used to fabricate an array of microscale GaAs diode switches. Basic fabrication route and the DC characterization procedure make use of those previously employed by us in \cite{shurakov2018ti}. The key difference is that the Ti/Pt/Au (50/20/20 nm) Schottky contact is formed next to alignment marks at the very beginning of the fabrication procedure within a lift-off process. The contact metallization is deposited through an opening with a diameter of 3.5 $\mu$m in 250 nm thick SiO$_2$ mask. Wafer surface is preliminary processed by O$_2$ plasma (50 W, 10 sccm, 15 mTorr) and wet etching in H$_2$O:H$_2$O$_2$:Ammonia solution 25$\%$ (80:8:2). We use GaAs wafers utilizing the n/n$^+$ (0.2/2 $\mu$m) sandwich implemented on a 350 $\mu$m thick SI-GaAs. The wafers are produced by the means of metalorganic chemical vapor deposition (MOCVD). The n and n$^+$ layers have the dopant concentrations of $4\times10^{17}$ and $5\times10^{18}$ cm$^{-3}$, respectively. 

\begin{figure}[!b]
\centering
\includegraphics[width=2.5in]{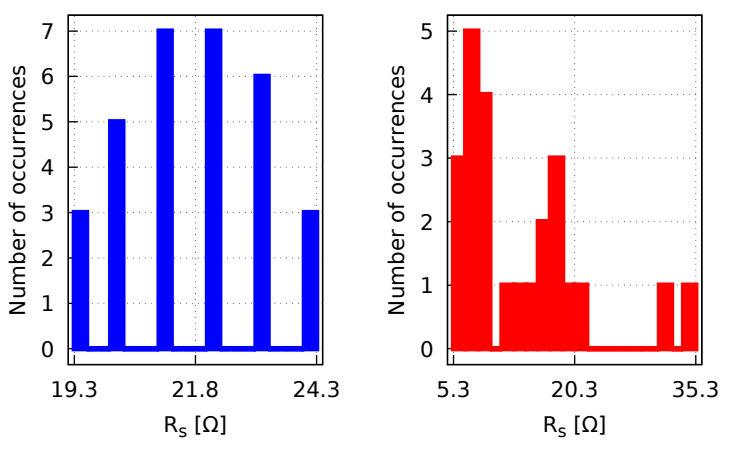}
\caption{Statistical data on the series resistance of Shottky diodes annealed at 350$^{\circ}$C (left panel) and 450$^{\circ}$C (right panel).}
\label{fig9}
\end{figure}

Once ohmic contacts are fabricated, the diodes are annealed at 350$^{\circ}$C. After performing a screening procedure, we anneal the diodes again. The latter routine is carried out twice: $T_{rta}$ is consequently set to 400 and 450$^{\circ}$C, the process duration is fixed to 60 s.

Data on the diodes’ series resistance ($R_s$) measured after the first and the third iterations of RTA is provided in figure \ref{fig9}. The statistics is obtained for an array of 32 diodes, and it yields the decrease of the $R_s$ mean value ($\bar R_s$) of 8.4~$\Omega$ in response to sweeping $T_{rta}$ from 350 to 450$^{\circ}$C. This corresponds to a 10\% increase in the Schottky barrier height, ideality factor is not noticeably affected by RTA and stays within 1.1–-1.2. Note that position-dependent resistance of the traces used for current supply of the array elements (2.2--3.2~$\Omega$) is not calibdrated out and may moderately affect statistics on intrinsic properties of the diodes. Homogeneity of the array elements is pronounced for $T_{rta}$ = 350$^{\circ}$C, the percentage of good diodes of 97$\%$ is obtained in this case. Elevating $T_{rta}$ upto 450$^{\circ}$C, we achieve better individual performance for 64.5\% of the diodes (29\% reveals the $R_s$ improvement greater than by a factor 2) at the cost of the 22\% yield reduction and increase of the diodes’ parameters spread by a factor of 5.6. Three local distributions with $\bar R_s$ of 7, 18 and 33 $\Omega$ are distinguished for $T_{rta}$ = 450$^{\circ}C$. To evaluate contribution of the ohmic contact formation process to the trends observed, we obtain KPFM images of the ``ladders'' for a number of diodes belonging to each of the three groups. The experiment reveals no noticeable difference of the corresponding n$^+$-GaAs surface potentials measured relative to the Au zero level which is interpreted as a lack of significant turbulence of the nitrogen flow during RTA. Furthermore, visual inspection of the array surface via optical microscope suggests that the observed negative effects are due to mild degradation of mutual ground traces used for current supply of the array elements. The fact that passive transmission line may noticeably compromise performance of an active electronic device once again proves importance of non-contact quality control in fabrication of a complex IC.

\section{Conclusion}
In this work, we develop a ladder-like layered ohmic contact to Si-doped GaAs with a dopant concentration of $5\times10^{18}$~cm$^{-3}$ and cross-study it via Kelvin probe force microscopy and grooved transmission line method. The ohmic contact metallization system is based on a Ni/Au/Ge/Ni/Au (5/45/20/5/100 nm) sandwich with an offset between Ni/Au/Ge and Ni/Au stacks of a few microns. Such a geometry is developed to enable measurement of the surface potentials of Ge and n$^+$-GaAs relative to that of Au acting as a reference level. We investigate the effect of rapid thermal annealing and measure contact resistivity as low as 0.15~$\mu \Omega \,$cm$^2$ resulting in only a 0.6~$\Omega$ of resistance for the contact area of 3$\times$3~$\mu$m$^2$. The lowest resistance is achieved for the ohmic contacts annealed at 450$^{\circ}$C. This temperature is also considered to be more efficient in case of thick semiconductor due to the observed resistivity distribution below the ohmic contact surface and is characterized by equalizing of the surface potentials of Ge and n$^+$-GaAs in the Au-Ge-n$^+$-GaAs ``ladder''. This finding is used to evaluate impact of the ohmic contact formation process on series resistance of a GaAs Schottky diode. We fabricate an array of 32 microscale GaAs diode switches which is inspired by our ongoing development of a millimeter wave intelligent reflective surface for 6G communications. Such a surface potentially comprises hundreds of fast and ultracompact diode switches and is used for beamforming in reflected light. The diodes must be identical and have a series resistance of a few ohms. We observe 3 local distributions of a series resistance with mean values of 7, 18 and 33 $\Omega$. Experiment reveals no noticeable difference of the surface potentials of n$^+$-GaAs measured relative to Au in the ``ladders'' for the diodes regardless their series resistances. We interpret this as a lack of significant turbulence of the nitrogen flow during the annealing process. Visual inspection of the array surface via optical microscope confirms high quality of the diodes and attributes the observed negative effects to degradation of current supply traces in the array. Thus, we propose that the surface potential mapping of the ``ladders'' can be used as the method of nondestructive quality control in fabrication of GaAs high-speed electronics.

\section*{Acknowledgements}
Section 1 is written under the support of the Basic Research Program of the National Research University Higher School of Economics. Sections 2--4 are written under the support of the Russian Science Foundation grant No. 22-79-10279, \url{https://rscf.ru/project/22-79-10279/}.

\bibliographystyle{ieeetr}
\bibliography{main}  

\end{document}